**Temperature Driven Structural Phase Transition in Tetragonal-Like BiFeO$_3$**

Wolter Siemons,[1] Michael D. Biegalski,[2] Joong Hee Nam,[1,3] and Hans M. Christen[1*]

[1] *Materials Science and Technology Division, Oak Ridge National Laboratory, Oak Ridge, TN 37831, USA*
[2] *Center for Nanophase Materials Science, Oak Ridge National Laboratory, Oak Ridge, TN 37831, USA*
[3] *Optic and Electronic Ceramics Division, Korea Institute of Ceramic Engineering and Technology (KICET), Seoul 153-801, Republic of Korea*

Abstract:

Highly-strained BiFeO$_3$ exhibits a "tetragonal-like, monoclinic" crystal structure found only in epitaxial films (with an out-of-plane lattice parameter exceeding the in-plane value by >20%). Previous work has shown that this phase is properly described as a M$_C$ monoclinic structure at room temperature [with a (010)$_{pc}$ symmetry plane, which contains the ferroelectric polarization]. Here we show detailed temperature-dependent x-ray diffraction data that evidence a structural phase transition at ~100 °C to a high-temperature M$_A$ phase ["tetragonal-like" but with a ($\bar{1}$10)$_{pc}$ symmetry plane]. These results indicate that the ferroelectric properties and domain structures of strained BiFeO$_3$ will be strongly temperature dependent.

[*]E-mail address: christenhm@ornl.gov

Bismuth ferrite [BiFeO$_3$, ("BFO")] is the only known material that exhibits both magnetic and ferroelectric order at room temperature and has thus received much attention.[1] While the multiferroicity of this perovskite has been the subject of much interest, the ferroelectric and structural properties by themselves are fascinating. In fact, the ferroelectric polarization, which in the bulk points along the pseudocubic [111]$_{pc}$ direction, rotates towards the surface normal (i.e. the [001]$_{pc}$ direction) under mild compressive strain[2,3] (here and in the following, the subscript "pc" is used to denote pseudocubic indices). This strain is imposed onto the film by its epitaxial relationship to a substrate such as SrTiO$_3$, with a = 3.905 Å (while for BFO, $a_{pc}^{bulk} \approx$ 3.96 Å). The crystal structure of such films is described as monoclinic, and deviates little from that of the bulk analog. Therefore, we refer to this phase as rhombohedral-like ("R-like").

A drastic change in crystal structure occurs when the compressive in-plane strain exceeds σ ≈ -4.5% (which is achieved, for example, by epitaxial growth onto LaAlO$_3$ ("LAO") substrates, $a_{pc}$ = 3.789 Å): the out-of-plane lattice parameter increases in step-like fashion by more than 20%, resulting in a monoclinically-distorted structure that resembles a highly-tetragonal unit cell ("T-like").[4-7] Interestingly, the change from the monoclinic R-like to the monoclinic T-like phase is accompanied by a significant symmetry change.[8,9] To fully appreciate these differences, we "construct" the two different monoclinic structures by subjecting a hypothetical tetragonal unit cell to a shear distortion perpendicular to the long axis (and thus, in the case of BFO films, parallel to the film/substrate interface plane). As illustrated in Fig. 1, a shear distortion in the [100]$_{pc}$ direction results in a monoclinic structure for which the mirror (or glide) plane is parallel to the (010)$_{pc}$ plane. Neumann's principle[10] dictates that the ferroelectric polarization is contained in this plane. Using the notation introduced by Vanderbilt and Cohen[11] we refer to this phase as M$_C$. In contrast, a so-called M$_A$ structure is obtained when the hypothetical tetragonal unit cell is subject to a shear distortion along the [110]$_{pc}$ direction (see Fig. 1); the ferroelectric polarization then lies within the ($\bar{1}$10)$_{pc}$ plane. As has recently been shown,[8,9] the strain-induced R-like–to–T-like transition at room temperature is in fact a M$_A$–to–M$_C$ symmetry change, with the corresponding necessary change of the polarization orientation.

Inspection of Fig. 1 shows that a doubled unit cell is obtained in the M$_A$ case. For simplicity, the unit cells shown correspond to the smallest ones compatible with the symmetry as evidenced by x-ray diffraction but ignore any further doubling due to octahedral tilts or antiferromagnetic spin order. The following arguments allow us to distinguish between M$_A$ and M$_C$ from x-ray reciprocal space maps (RSMs). Here we consider films having a domain structure in which the film's and the substrate's (00L) planes are parallel. This, as we discuss below, is the case in our samples. In this case, RSMs through the family of peaks corresponding to the monoclinically-indexed {H0L} and {HHL} planes show a triplet and a doublet, respectively.[12,13] Considering that in the case of the M$_A$ structure the monoclinic *a* and *b* axes are rotated by 45° with respect to the pseudocubic (perovskite) axes, comparing the RSMs taken in two pseudocubic orientation (see Table I) immediately distinguishes between M$_A$ and M$_C$.

M$_A$-to-M$_C$ transitions are not unique to BFO; in fact, they have first been observed in Pb-based solid-solution perovskites, where they occur not directly as a consequence of strain but are induced by electric fields or changes in temperature.[14] Interestingly, a temperature-induced symmetry change from the room-temperature M$_A$ phase to a higher-temperature M$_C$ structure has recently been observed for low-strain (i.e. R-like) BFO films.[15] Note that in this R-like high-temperature $M_C$ phase, the authors find b>a, which would intuitively not be expected for a ferroelectric phase, and in fact, the $M_A$-to-$M_C$ transition is postulated to correspond to the ferroelectric-paraelectric phase transition. Additionally, the symmetry change is associated with an abrupt ~ 1% reduction in the c-axis lattice parameter. In contrast, we have recently studied the temperature-dependence of the c-axis lattice parameter of T-like BFO films on $LaAlO_3$ substrates[16] and observe a monotonic behavior between 100°C and the decomposition of the film at 750°C. In other words, the T-like distortion (i.e. the largely enhanced c-axis lattice parameter) is essentially independent of temperature for highly-strained BFO. Note that BFO exhibits a lower coefficient of thermal expansion ($\alpha_{BFO} \approx 0.6 \cdot 10^{-5}$/K)[16] than LAO ($\alpha_{LAO} \approx 1.1 \cdot 10^{-5}$/K),[17] and therefore the film's c-axis lattice parameter slightly contracts upon heating.

In this study, we pay closer attention to the monoclinic distortion within the T-like phase as a function of temperature, and show that it indeed changes from a room-temperature $M_C$ structure (T-like) to a higher-temperature $M_A$ structure (also T-like).

The structural properties are studied on a sample grown from a BFO sintered target with 10% excess Bi by pulsed laser deposition (PLD) with a KrF excimer laser (248 nm) at 5 Hz. During deposition a LAO substrate was kept at a temperature of 700 °C with 25 mTorr oxygen background pressure, resulting in a deposition rate of approximately 0.2 Å/pulse. The film analyzed in this work was approximately 300 nm thick.

X-ray diffraction measurements were performed on a PANalytical X'Pert thin film diffractometer with Cu $K_\alpha$ radiation equipped with an Anton Paar hot stage. Diffraction patterns were obtained at 25 °C intervals between 25 and 175 °C. In Figure 2a we show the diffraction pattern along the crystallographic $[00l]_{pc}$ direction at 25 and 175 °C. Besides the expected BFO and LAO peaks there is a small contribution from an unassigned additional epitaxial phase that we discuss below (labeled with a star symbol), with is observed both at the $001_{pc}$ and the $002_{pc}$ locations. Peaks corresponding to excess bismuth oxides are labeled. The additional peak near 44° results from the sample hot stage and is absent in measurements performed on the standard sample holder. Therefore, the data shows that our films are phase-pure with the exception of excess bismuth oxides and the temperature-dependent observation of the above-mentioned phase indicated by a star. In particular, there is no presence of the R-like BFO (which would be visible near 45.7°) or the recently-observed intermediary phase (near 43.4°).[18]

In figure 2b we focus on the evolution of the $004_{pc}$ peak as a function of temperature. A clear phase transition is observed at 100 °C: the peaks of the low-temperature phase (near 82.5°) diminish in intensity as the high-temperature phase appears. Intrinsic peak broadening makes it impossible to observe this splitting at lower angles such as those investigated previously.[16] However, it is clearly seen in Fig. 2b that higher-temperature peaks shift towards higher angles (i.e. smaller lattice parameters) with increasing temperature, consistent with our previous observations of a lower thermal expansion of BFO than of LAO.

To understand the nature of this phase transition near 100°C, we use RSMs through the $002_{pc}$, $103_{pc}$, and $113_{pc}$ reflections. At all temperatures, the $002_{pc}$ maps show only one peak for the film, confirming that the film's (00L)-planes are parallel to those of the substrate, as discussed above.

The results of the maps containing the $103_{pc}$ and $113_{pc}$ reflections shown in figures 3a and b. At room temperature (Fig. 3a) we find a triplet in the 103 map and a doublet for 113, which, by comparing to Table I, reconfirms the previous observation[8,9] of an $M_C$ phase. From the peak positions, we find that a/b = 1.019(2), c = 4.67(2)Å, and β = 88.1(3)°. However, as the sample is heated above the phase transition temperature to 175 °C, we find that the $103_{pc}$ peak now shows a doublet and the $113_{pc}$ peak a triplet. Inspection of Table I immediately indicates that the film at high temperature exhibits $M_A$ monoclinic phase, where the monoclinic distortion is along the $[110]_{pc}$ direction. Confirming the consistency of this interpretation, we quantitatively inspect the splitting of the peaks in the normal ($[001]_{pc}$) direction of the RSMs: At 25°C, we observe $\{q_Z(\bar{1}13_m)- q_Z(113_m)\} = \{q_Z(\bar{1}03_m)- q_Z(103_m)\}$ as expected. At 175°C, the $113_{pc}$ map corresponds to a higher-order monoclinic diffraction, and therefore $\{q_Z(\bar{1}13_m)- q_Z(113_m)\} = \frac{1}{2} \{q_Z(\bar{2}03_m)- q_Z(203_m)\}$. Thus, our data can reliably be interpreted as monoclinic $M_A$ with a/b = 1.002(2), c = 4.67(2)Å, and β = 88.1(3)°. Note that the deviation of β from 90° is much larger in both T-like phases than it is in the R-like structure where β ≈ 89.5° is reported.[2,3] In other words, the T-like phase has a large c/a ratio but is otherwise further from tetragonal than the R-like structure. We also note that two recent studies report results from x-ray diffraction and Raman spectroscopy [19] or neutron scattering [20] that are consistent with our observation of a structural phase transition near 100°C, and those findings can now be attributed to the $M_C$–to–$M_A$ symmetry change.

The temperature-induced change from one monoclinic structure to a different one has profound consequences for the ferroelectric properties of BFO films but is not strikingly different from what can be expected based on the current knowledge of BFO and related materials. As mentioned above, such transitions occur in Pb-based solid-solution ferroelectrics and in R-like BFO. In addition, previous calculations[9] have shown the energetic proximity of several different crystalline structures within T-like BFO. There, it was also pointed out that the transformation from the $M_C$ to the $M_A$ phase cannot be continuous – it requires the co-existence of an intermediate phase or a co-existence of two competing phases, and this co-existence is clearly observed here (Fig 1b). Therefore, in the intermediate temperature range (near 100 °C), the ferroelectric and structural properties will be highly sensitive to in-plane electric or elastic stimuli, as the projection of the ferroelectric polarization onto the in-plane direction differs between the co-existing $M_A$ and $M_C$. Thus, interesting new piezoelectric effects are expected, and there is hope that slight chemical modifications (doping) might lower the structural phase transition to room-temperature, further increasing the practical utility of this material.

Finally we note that the high temperature $M_A$ monoclinic phase approaches a higher symmetry than the room-temperature $M_C$ [b/a closer to unity, and thus also $b_{pc} \approx a_{pc}$ with $\angle(a_{pc}, b_{pc}) \approx 90°$]. While changes of the ferroelectric and ferroelastic properties across the transition are still being investigated, these observations shed light on the formation of the domain structures in T-like BFO films. In fact, the high-temperature $M_A$ phase (which we assume to be present at the growth temperature) exhibits a nearly-square in-plane lattice, as required for good epitaxial match. This is not the case for the room-temperature $M_C$ structure [with b/a = 1.019(2) as discussed above]. Clearly this temperature-reversible change has to occur without the breaking of chemical bonds and will thus locally lead to large elastic strains. However, such strains cannot be fully accommodated within a monoclinic structure of the observed texture [i.e. having the (00L)- planes of the film parallel to those of the substrate]. Therefore, it is not surprising that an additional phase (indicated by a star symbol in Fig. 2a) occurs at low temperature.

To summarize, we observe a temperature-driven phase transition in tetragonal-like BFO from a monoclinic $M_C$ phase at room temperature to a different (but still T-like) monoclinic phase of $M_A$ symmetry. This will have significant consequences for the ferroelectric and ferroelastic properties of T-like BFO, as the symmetry constrains the in-plane component of the polarization to different orientations in these two phases (along $[100]_{pc}$ at room temperature but along $[110]_{pc}$ above 100°C). The high-temperature $M_A$ phase approaches a higher symmetry in its in-plane arrangement than its room-temperature $M_C$ counterpart, with a nearly square lattice observed. Future work is needed to study the effect of this structural phase transition on optical, ferroelectric, and magnetic properties.

Acknowledgments
W.S. and H.M.C. acknowledge support by the U.S. Department of Energy, Office of Basic Energy Sciences, Materials Sciences and Engineering Division. X-ray diffraction (M.D.B.) was supported by the Center for Nanophase Materials Sciences (CNMS), which is sponsored by the Office of Basic Energy Sciences, US Department of Energy. J.H.N. was supported by the Republic of Korea, Ministry of Knowledge and Economy, Visiting Scientists Program, under IAN:16B642601, with the US Department of Energy.

Tables

**Table I.** Monoclinic diffraction peaks visible in the pseudocubically indexed RSMs. Observations of a doublet or triplet immediately distinguishes between $M_A$ and $M_C$.

| RSM | $M_A$ | $M_C$ |
|---|---|---|
| $103_{pc}$ | $\bar{1}13, \bar{1}\bar{1}3$ | $\bar{1}03$ |
|  | $113, 1\bar{1}3$ | $013, 0\bar{1}3$ |
|  |  | $103$ |
| $113_{pc}$ | $\bar{2}03$ | $\bar{1}13, \bar{1}\bar{1}3$ |
|  | $023, 0\bar{2}3$ | $113, 1\bar{1}3$ |
|  | $203$ |  |

Figures

Figure 1:

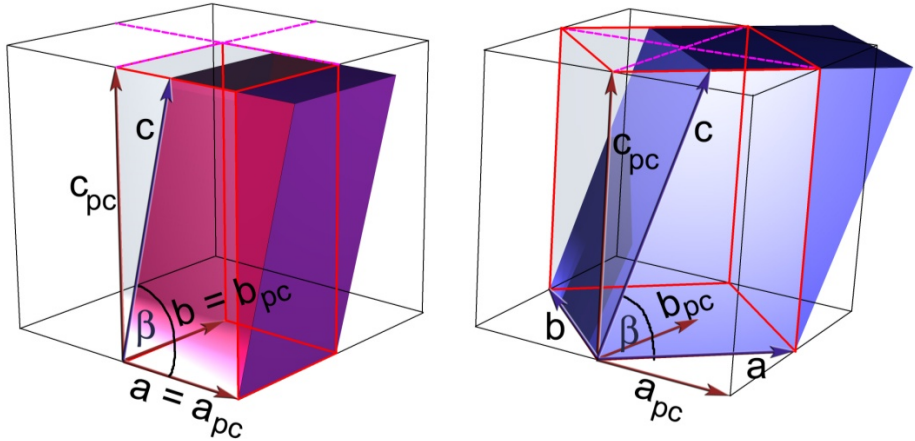

Figure 1. (Color online) Schematic drawing of monoclinic unit cells. (a) depicts a $M_C$ structure, in which a ∥ $a_{pc}$ and b ∥ $b_{pc}$ (where a and b are the proper monoclinic axes), and which can be seen as resulting from a shear distortion along the $[100]_{pc}$ direction. (b) shows the result of a shear distortion along $[110]_{pc}$, requiring a doubled unit cell with the a and b axes rotated by 45° with respect to their pseudocubic counterparts. The smallest unit cells consistent with these shear distortions are shown and used to index the data in this paper, ignoring additional doubling of the cells due to antiferromagnetism and octahedral tilts.

Figure 2:

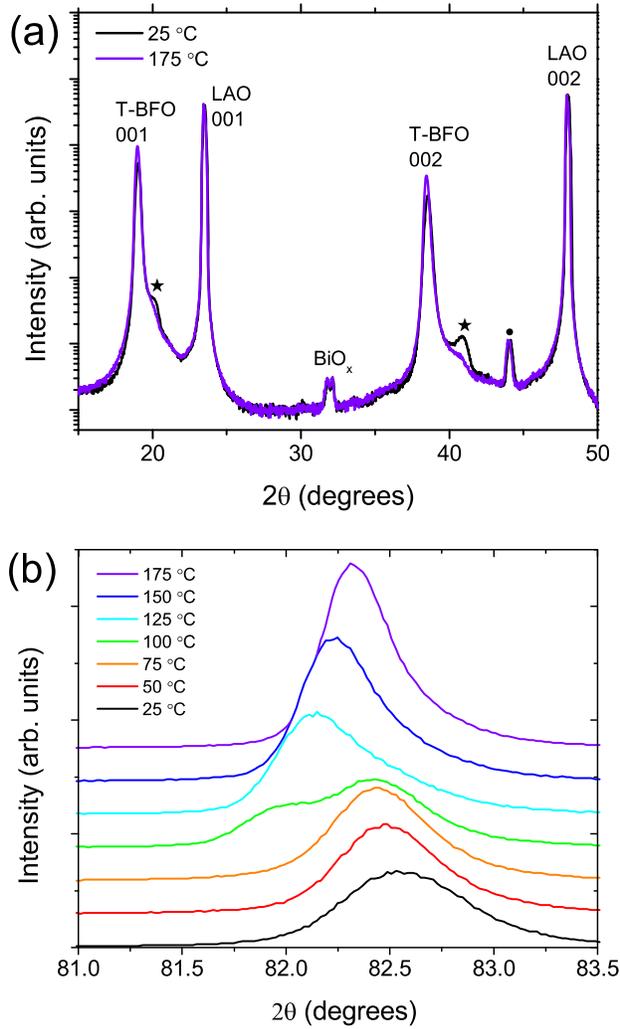

Figure 2. (Color online) X-ray θ-2θ scans for T-like BFO. (a): Scans through the $001_{pc}$ and $002_{pc}$ peaks of the BFO film and the LAO substrate, at 25°C and 175°C. The peak labeled with a dot (●) results from the hot-stage. The peaks labeled with a star (★) are epitaxial peaks of a secondary room-temperature phase. Also indicated are peaks corresponding to excess bismuth oxide. (b): expanded view through the film's $004_{pc}$ peak at 25°C intervals upon heating. Curves are displaced vertically for clarity. At 100 °C, the co-existence of two T-like phases is observed.

Figure 3:

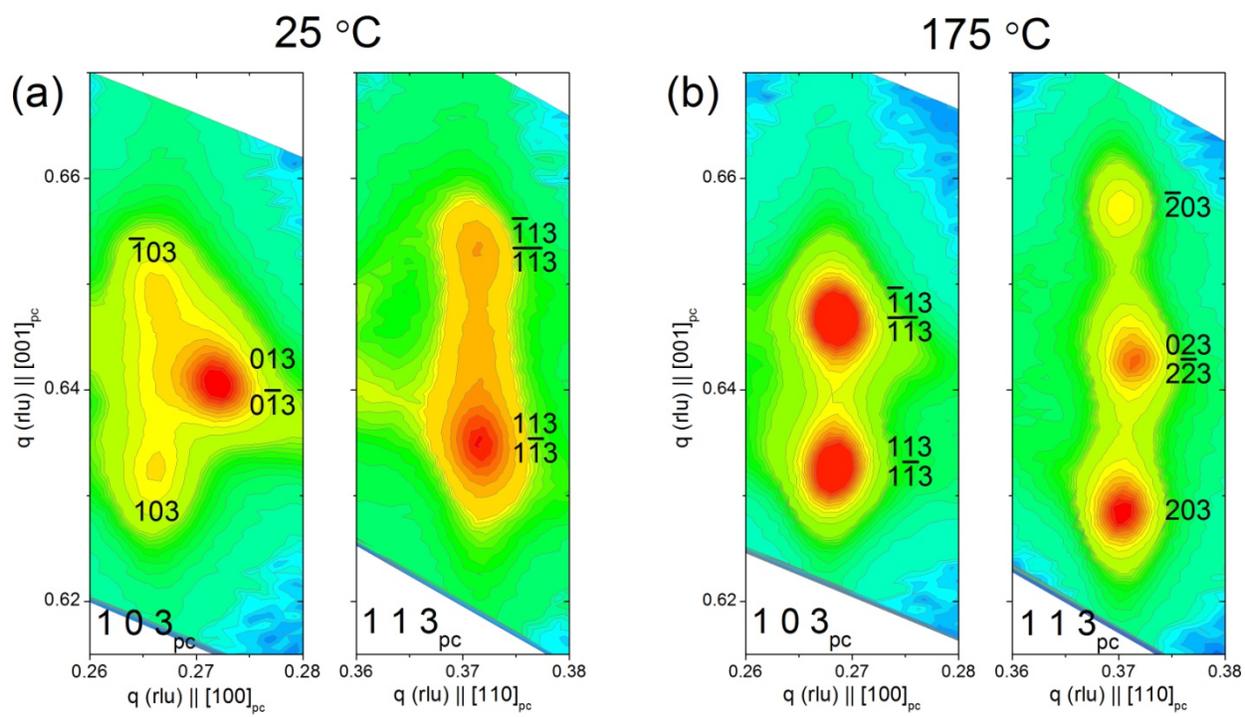

Figure 3. (Color online) RSMs taken through the BFO peaks in the pseudocubic <103>$_{pc}$ and <113>$_{pc}$ directions, at 25 °C (a) and 175 °C (b). Monoclinic peak positions are indexed in the unit cells drawn in Fig. 1(a) and (b), respectively.